# Bridging microscopic nonlinear polarizations toward far-field second harmonic radiation

KYUNGWAN YOO,[1] SIMON F. BECKER,[2] MARTIN SILIES,[2] SUNKYU YU,[1,*]
CHRISTOPH LIENAU,[2] AND NAMKYOO PARK[1,*]

[1]*Photonic Systems Laboratory, Department of Electrical and Computer Engineering, Seoul National University, Seoul 08826, Korea.*
[2]*Institut für Physik and Center of Interface Science, Carl von Ossietzky Universität, D-26111 Oldenburg, Germany.*
*\*nkpark@snu.ac.kr, \*skyu.photon@gmail.com*

**Abstract:** Since the first observation of second harmonic generation (SHG), there have been extensive studies on this nonlinear phenomenon not only to clarify its physical origin but also to realize unconventional functionalities. Nonetheless, a widely accepted model of SHG with rigorous experimental verification that describes the contributions of different underlying microscopic mechanisms is still under debate. Here, we examine second harmonic far-field radiation patterns over a wide angle from metallic structures with different resonances, to reveal the structure-dependent contributions from distinct nonlinear polarizations. By comparing the measured SHG radiation patterns of 82 antennas with different SHG models, we demonstrate the critical role of the surface-parallel and bulk nonlinear polarizations in the far-field SHG patterns, and thus show that the common belief of the dominant contribution of the surface-normal component in SHG should be corrected. A virtual multi-resonator SHG model inside a single physical resonator is introduced to explain and control the interplay between different nonlinear polarizations and their structure-dependent excitations. Our findings offer a new strategy for the design of highly efficient and directional nonlinear metamaterials.

## 1. Introduction

Second harmonic generation (SHG) is one of the most salient phenomena in nonlinear optics, both in terms of its symmetry-dependent manifestation and unique applications [1,2]. Since the first observation of SHG from quartz [3], the underlying physical mechanisms and their connection to material and/or structural symmetry have been studied intensively in relation to inversion symmetry [4,5], surface and bulk contributions [6-8], hydrodynamical electron pressure [9-11], plasmon- or exciton-resonant enhancement [12-17], and geometrical selection rules [18]. The underlying physics of SHG, including frequency conversion and a strong symmetry dependence, have also facilitated diverse applications such as spin-wavelength multiplexed holography [19], exciton spectroscopy [20], an all-optical diode [21], a tunable optical transistor [22], and biomedical target imaging [23,24]. In particular, compared to the initial studies that considered relatively simple geometries of flat films [4,25] and spheres [8,26], recent SHG studies have focused on the combination of multiple metallic elements to emphasize inter-particle couplings for larger degrees of freedom in the manipulation and enhancement of far-field radiation [27-31].

While this trend of multi-element SHG control is a logical path toward realizing SHG applications, in-depth explorations of the microscopic mechanism of SHG at the fundamental level and its far-field conversion based on individual nonlinear polarization components are surprisingly rare [30,32,33], even at the single SHG particle/element level. While the excitation of SHG in a centrosymmetric material is theoretically well explained by the field gradient in the bulk (bulk nonlinear susceptibility) and symmetry breaking at the surface

(surface-parallel- and normal components of the nonlinear susceptibilities) [2,8], there exist order-of-magnitude discrepancies between the reported susceptibility values [7,8,34]. At present, the common practice is to assume that the surface-normal *susceptibility* component is always dominant [14,15,35]: which is obtained from far-field SHG measurements, yet made for limited geometries of metals (a flat film [7] and a sphere of subwavelength size [8]). Moreover, while a large nonlinear *susceptibility* does not guarantee an efficient far-field *SHG radiation*, there often exists confusion between those two notions [13,27,28,36]. As demonstrated in a flat metal film [7], dominant surface-*parallel* SHG *radiation* from a large surface-*normal* SHG *susceptibility* is possible. Overall, with only a few datasets available for measured susceptibility values under restricted conditions, together with the lack of understanding of a relation that bridges the microscopic susceptibility and the far-field SHG radiation, it is conventional to take the surface-normal susceptibility component as dominant dynamics both in the susceptibility level and far-field radiation, irrespective of the geometries of metallic particles [13-15,27,28,35,36]. This lack of an established SHG model at current stage thus constitutes a significant bottleneck in the development of future SHG devices and also raises the following question: Is it possible to control SHG in the far-field by manipulating individual polarization sources?

In this paper, we dispute the dominant contribution of the surface-normal component, by revealing the critical role of the surface-parallel and bulk nonlinear polarizations in SHG both at the susceptibility and far-field radiation levels. By conducting angle-resolved *k*-space measurement [37,38] over a wide angle (± 67°), for SHG radiation patterns from 82 antennas with subwavelength- ($\lambda_0/7$) to wavelength-scale ($\lambda_0$) footprints, we successfully extract the geometry-dependent variation in the far-field SHG radiation. By comparing the experiment to the numerical results obtained for different nonlinear susceptibility models [7,8,11,34,39], we then show that the susceptibility parameters based on a hydrodynamic model (HDM) [11] provides the best fit. Most critically, we also reveal that the contributions of the surface-parallel or bulk components to the SHG radiation are in general comparable to or even larger than the contribution of the surface-normal component, in contrast to previous reports emphasizing the dominant role of the surface-normal component [13-15,27,28,35,36]. Finally, we also show that *each* polarization component (bulk, surface-parallel, and surface-normal) exhibits different couplings to the far-field radiation with a distinct dipole direction and distribution, and thus can be treated as a virtual resonator with its resonance property depends on different geometrical parameters (width, height, sharpness) of the antenna. By clarifying the significance of bulk susceptibility and surface-parallel SHG susceptibility, and controlling the SHG *radiation pattern* with the characteristic contribution from each nonlinear polarization component, our result extends the result of O'brien *et al*. [32], where the SHG conversion *efficiency* was analyzed in view of the nonlinear polarization distributions, yet with a model biased toward a surface-normal SHG susceptibility [7]. By clarifying the role and contribution of each nonlinear polarization component and thereby bridging the microscopic tensor to the far-field radiation, our findings imply the need for new design strategies for highly efficient and directive SHG devices, especially utilizing metamaterials with complex geometries, layouts, and inter-particle couplings.

## 2. Result and discussion

To examine microscopic nonlinear polarizations in SHG, we employ gold antennas [Fig. 1(a)] fabricated on top of a glass substrate with electron beam lithography, a lift-off process, and an annealing technique (see Section 1 in the Supplement 1 for details). With the excitation of femtosecond pulses (polarized in the *x*-direction, 12 fs, average power of 9.5 mW; laser and SHG spectra are given in Fig. S1 in the Supplement 1), the linear resonances of these antennas are predominantly determined by the width of the antenna $w$ [132 ~ 964 nm, 10 nm interval, see Fig. 1(b)], with negligible dependences on the thickness $t$ (35 nm) and depth $d$ (5 μm) [Fig. 1(c)]. In the experiment, generated SHG signals from the antennas are measured in the back focal plane of an oil immersion objective lens (NA = 1.65), which provides the Fourier transformed (*k*-space) radiation pattern [37,38] (see Section 1 and Fig. S2 in the Supplement 1). With index of glass substrate (*n* = 1.79), the numerical aperture of this objective lens (NA = 1.65) allows to collect the SHG signals over a wide angle (-67° < $\theta$ < 67°) in transmission direction [see inset of Fig. 1(c)].

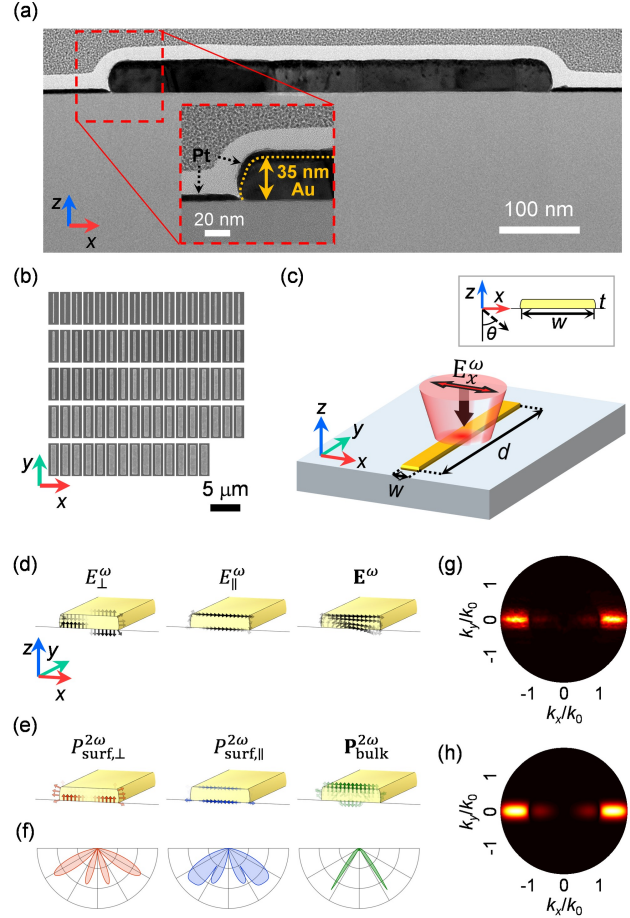

**Fig. 1**. Fabricated antenna samples and *k*-space SHG imaging of a single antenna. (a) TEM cross-sectional image of a fabricated antenna. An additional Pt layer (5 nm) was deposited after the SHG measurements for SEM imaging. The yellow dotted line in the inset indicates the estimated gold boundary. (b) SEM images of the fabricated 82 gold antennas. The distribution of each antenna on the wafer can be found in Fig. S6 in the Supplement 1. (c) Schematic of the SHG excitation. A gold antenna on a glass substrate (*n* = 1.79) is excited by an *x*-polarized, normal-incident Gaussian beam. SHG signals are collected in the transmission direction. (d),(e) Calculated (d) electric field components at the fundamental frequency and (e) nonlinear polarization components at the second harmonic frequency for an antenna width *w* = 420 nm. (f) Angle-resolved SHG radiation patterns calculated from each nonlinear polarization component of (e). (g) Measured and (h) calculated SHG intensity in *k*-space from a single gold antenna (*w* = 215 nm).

For the numerical analysis, first, the finite element method (FEM) calculation at the fundamental frequency is conducted [Figs. 1(d)], by assuming Gaussian wave illumination (beam waist of 893 nm) for the 3D antenna structures, geometry of which is obtained from a transmission electron microscopy (TEM) image [Fig. 1(a)]. The fundamental wavelength ($\lambda$ =

888 nm) used in this calculation is deduced from the peak wavelength ($\lambda = 444$ nm) of the measured SHG spectra (see Fig. S1 in the Supplement 1). For the calculation of SHG radiation, each nonlinear polarization component (surface-normal [$P^{2\omega}_{\text{surf},\perp} = \varepsilon_0 \chi_{\perp\perp\perp} E^\omega_\perp E^\omega_\perp$], surface-parallel [$P^{2\omega}_{\text{surf},\parallel} = \varepsilon_0 \chi_{\parallel\parallel\perp} E^\omega_\parallel E^\omega_\perp$], and bulk [$\mathbf{P}^{2\omega}_{\text{bulk}} = \varepsilon_0 \gamma \nabla(\mathbf{E}^\omega \cdot \mathbf{E}^\omega) + \varepsilon_0 \delta'(\mathbf{E}^\omega \cdot \nabla)\mathbf{E}^\omega$]) [Fig. 1(e)] is obtained by using the FEM-calculated fundamental fields ($\mathbf{E}^\omega$), and then used as a source for the FEM calculation at the second-harmonic frequency $2\omega$ (see Section 1 in the Supplement 1 for details). Figures 1(g) and 1(h) show excellent agreement between the experimental and theoretical SHG signals in $k$-space for a single antenna with $w = 215$ nm, in which the SHG signals are polarized in the $x$-direction (see Fig. S3 in the Supplement 1) and well confined near $k_y = 0$ (see Fig. S4 in the Supplement 1 for more data with other antenna widths). It is emphasized that obtained far-field SHG from each nonlinear polarization component exhibits strongly distinctive SHG radiation patterns [Fig. 1(f)].

More detailed studies on the SHG $k$-space radiation patterns have been carried out for a larger set of antennas, to excite different resonance modes at the fundamental frequency and to generate corresponding SHG radiation patterns. First, Fig. 2(a) shows the experimentally measured $k_x$-projected SHG radiation intensity as a function of the antenna width $w = 132 \sim 964$ nm, which exhibits an excellent fit with the numerical results obtained from the susceptibility parameters based on HDM [Fig. 2(b), see Section 2 in the Supplement 1]. Next, we also display in Fig 2(c)-2(e) the numerically obtained SHG radiation intensity, which used the *individual* nonlinear polarization components ($P^{2\omega}_{\text{surf},\perp}$, $P^{2\omega}_{\text{surf},\parallel}$, and $\mathbf{P}^{2\omega}_{\text{bulk}}$) only in the calculation. From all the figures 2(a)-2(e), it is evident that fitting with a single nonlinear polarization component [Figs. 2(c)-2(e)] does not provide a complete description of the experimental result [Fig. 2(a)]. Most critically, the surface-normal component $P^{2\omega}_{\text{surf},\perp}$ *alone* [Fig. 2(c)] does not result in a reasonable fit with the experimental data [Fig. 2(a)], in sharp contrast to common belief of the dominant contribution of the surface-normal component [13-15,27,28,35,36]. This is in line with the observation that, numerical results obtained for other representative models [Figs. 2(g)-2(j), see Section 3 in the Supplement 1 for details], of relatively larger $P^{2\omega}_{\text{surf},\perp}$ than other components, also fail to reproduce the measured SHG data. Only the HDM with all susceptibility components [Fig. 2(b)] faithfully reproduces the experimental results over a wide range of antenna geometries and radiation angles. We attribute the relative strength of the nonlinear susceptibility of gold ($|\chi_{\parallel\parallel\perp}| \sim |\chi_{\perp\perp\perp}|$ for the HDM, in contrast to $|\chi_{\parallel\parallel\perp}| \ll |\chi_{\perp\perp\perp}|$ for other models) as the origin of the observed discrepancies between the fitting models. The various nonlinear susceptibility parameters ($\chi_{\perp\perp\perp}$, $\chi_{\parallel\parallel\perp}$, $\gamma$, $\delta'$, $\chi_{\perp\parallel\parallel}$) from the models, converted to so-called conventional definition [7] that are used in the calculations, can be found in Section 3 of the Supplementary 1.

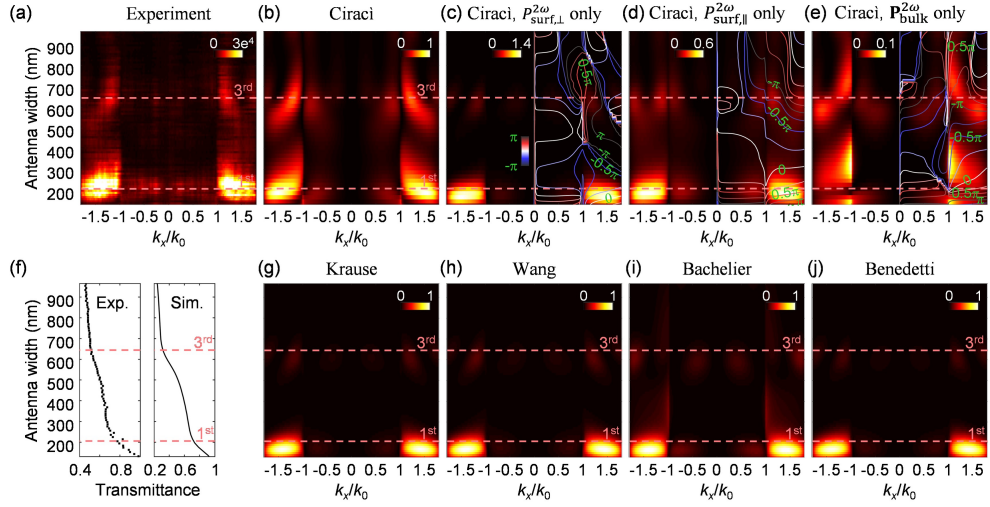

**Fig. 2.** (a) Measured and (b) HDM-calculated $k$-space-resolved SHG intensity (a.u.) as a function of antenna width. (c)-(e) SHG contributions from individual nonlinear polarization components: (c) surface-normal, (d) surface-parallel, and (e) bulk. The contour lines in (c)-(e) show the relative phases of the SHG in the far-field. The SHG intensities of (c)-(e) are normalized to the maximum value of (b). (f) Measured (left) and simulated (right) transmittances at the fundamental wavelength ($\lambda$ = 888 nm) as a function of antenna width. The red dotted lines correspond to antennas at the fundamental resonances: widths of 205 nm (1st) and 645 nm (3rd). (g)-(j) $k$-space-resolved SHG intensity (a.u.) calculated from established SHG models from (g) Krause [34], (h) Wang [7], (i) Bachelier [8], and (j) Benedetti [39].

Meanwhile, we also emphasize the antenna-width dependent variation in the SHG radiation pattern. It is observed that each nonlinear polarization component [Figs. 2(c)-2(e)] provides strongly distinct radiation *intensity* (color map) and *phase* (contour lines) patterns depending on the antenna width. This result highlights that observed SHG in the far-field is not a simple *intensity* sum, but a *phasor* sum of far-field SHG from different nonlinear polarization components, which again strongly depend on the geometry of the SHG element (here, as simple as the width of a linear antenna). It is worth to note that, the strong SHG signals for antenna widths of approximately $w$ = 205 nm and 645 nm [Fig. 2(a) and 2(b)] are in good agreement with the resonance widths of the antenna measured at the fundamental frequency [red dotted line in Fig. 2(f)], confirming the strong correlation between the SHG intensity and resonance at the fundamental frequency [27].

To better explain the observed SHG dependence on the nonlinear polarizations and geometrical parameters, in Fig. 3, we show the formation of individual nonlinear polarization components and their couplings to the radiation fields. First, upon the excitation of the antenna at the fundamental frequency, Figs. 3(a)-3(c) shows the FEM-obtained spatial distribution of the electric field inside the antenna [$\mathbf{E}^\omega$, Fig. 3(a)], in addition to the surface-parallel electric fields [$E_\parallel^\omega \sim E_x^\omega$, Fig. 3(b)] and surface-normal electric fields [$E_\perp^\omega \sim \rho_{surf}^\omega$, Fig. 3(c)] at the bottom surface, exhibiting distinct nodal distributions. With these fundamental field distributions, two major underlying mechanisms are considered to explain SHG radiation. First, as illustrated in Figs. 3(d)-3(f), it is noted that the distribution of each nonlinear polarization component $\mathbf{P}^{2\omega}$ is governed by the spatial profile of each constituting vector component of the electric field $\mathbf{E}^\omega$ (see Section 2 in the Supplement 1) as follows: $P_{surf,\perp}^{2\omega} \sim |E_\perp^\omega|^2$, prominent near the antenna corner because $E_\perp^\omega$ is the strongest in the high-surface-charge-density region [Fig. 3(d)]; $P_{surf,\parallel}^{2\omega} \sim |E_\parallel^\omega E_\perp^\omega|$, pronounced around the corner with $E_\perp^\omega$ but also spread out with an evenly distributed $E_\parallel^\omega$ [Fig. 3(e)]; and $\mathbf{P}_{bulk}^{2\omega} \sim |E_\parallel^\omega|^2$ in relation to the magnetic Lorentz force ($\mathbf{E}^\omega \times \mathbf{B}^\omega$, see Section 2 in the Supplement 1) showing a smooth

distribution throughout the antenna [Fig. 3(f)]. Second, each nonlinear polarization component provides a unique contribution to the far-field SHG couplings, not only from their distinct *distributions* for $P^{2\omega}_{\text{surf},\perp}$, $P^{2\omega}_{\text{surf},\parallel}$, and $\mathbf{P}^{2\omega}_{\text{bulk}}$ as evident in Figs. 3(d)-3(f) but also from their differences in *local dipole directions* [Figs. 3(g)-3(i)]. The polarization *distribution* and dipole *directionality* thus together determine the near- and far-field radiation from each polarization source [Figs. 3(g)-3(i)] and eventually the total SHG radiation.

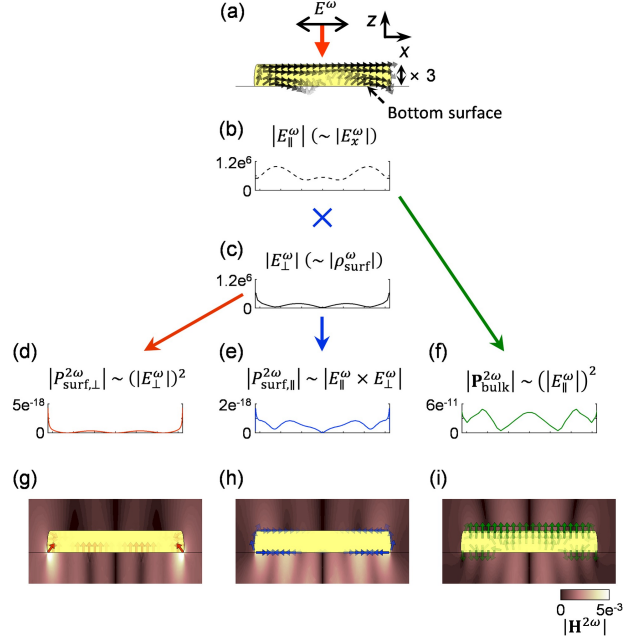

**Fig. 3**. Mechanisms of SHG radiation in terms of each nonlinear polarization component. (a) Induced fundamental electric field (black arrows) at the 3$^{\text{rd}}$-order resonance of the antenna. (b),(c) Spatial distributions of the (b) surface-parallel and (c) surface-normal fundamental electric fields at the bottom surface. (d)-(f) Spatial distribution of each nonlinear polarization component at the bottom surface, derived from the fundamental electric field in (a). (g)-(i) Calculated nonlinear polarization components (colored arrows) and corresponding SHG magnetic field amplitudes. Images of (a) and (g)-(i) are enlarged 3 times in the $z$-direction for clarity.

With the polarization components virtually acting as a set of three resonators for $P^{2\omega}_{\text{surf},\perp}$, $P^{2\omega}_{\text{surf},\parallel}$, and $\mathbf{P}^{2\omega}_{\text{bulk}}$ in one metallic antenna [Figs. 3(d)-3(f)], we also note that it is possible to manipulate the strength of each nonlinear polarization component in the far-field, by adjusting corresponding parameters in the geometry of the metallic antenna (*e.g.,* edge shape, width, and thickness; for details, see Section 4 in the Supplement 1).

With the understanding that each nonlinear polarization component $P^{2\omega}_{\text{surf},\perp}$, $P^{2\omega}_{\text{surf},\parallel}$, and $\mathbf{P}^{2\omega}_{\text{bulk}}$ offers distinctive contributions, it will be interesting to compare their *far-field intensity and angular distributions*. Figures 4(a)-4(d) show the *total* SHG radiation for different representative antenna widths, exhibiting excellent agreement between the experimental [dotted in Figs. 4(a)-4(d)] and numerical [lines in Figs. 4(a)-4(d)] results. Figures 4(e)-4(h) also display the SHG radiation for $P^{2\omega}_{\text{surf},\perp}$ (red), $P^{2\omega}_{\text{surf},\parallel}$ (blue), and $\mathbf{P}^{2\omega}_{\text{bulk}}$ (green) derived from the corresponding individual nonlinear polarizations [Figs. 4(i)-4(l)]. In general, it is found that $P^{2\omega}_{\text{surf},\parallel}$ (blue shade) provides comparable [Figs. 4(e) and 4(g)], or even larger [Figs. 4(f) and 4(h)] contributions to the far-field SHG pattern compared with $P^{2\omega}_{\text{surf},\perp}$ (red shade). Moreover, the angular distributions of each polarization component are heavily dependent on the antenna geometry via the resonance condition.

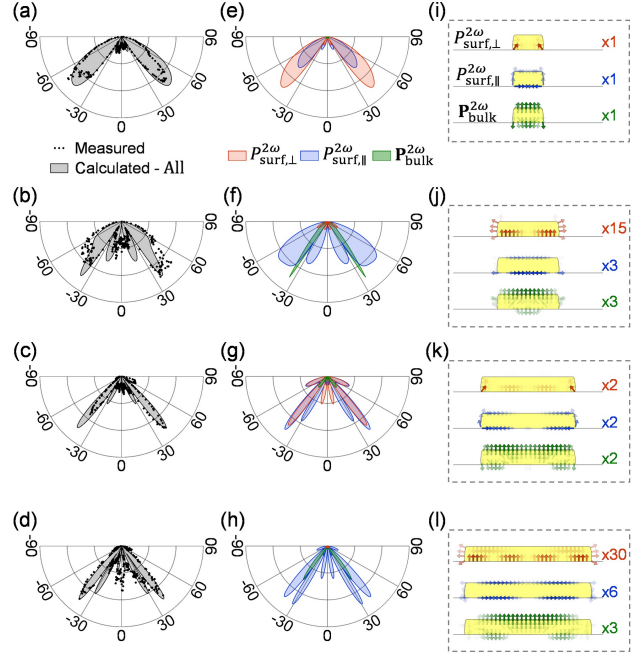

**Fig. 4**. SHG coupling to the far-field in terms of different resonance conditions and individual nonlinear polarization components. (a)-(d) Measured (black dots) and calculated (black line with shading) SHG radiation patterns for antenna widths of $w$ = 205 (a), 420 (b), 645 (c), and 870 (d) nm. (e)-(h) SHG radiation patterns from each nonlinear polarization component (the red, blue, and green lines with shading correspond to the results from the $P^{2\omega}_{surf,\perp}$, $P^{2\omega}_{surf,\parallel}$, and $\mathbf{P}^{2\omega}_{bulk}$ nonlinear polarizations, respectively). (i)-(l) Calculated nonlinear polarizations [using the same colors and the same set of antenna widths as in (e)-(h)]. The numbers in (i)-(l) represent normalization scales. As in Fig. 3, the images of (i)-(l) are enlarged 3 times in the z-direction for clarity.

With the distinct far-field radiation patterns of each polarization component under different conditions, even at the level of a single metallic element, this result implies that a proper consideration of all nonlinear polarization components is critical, especially for SHG devices with complex geometries and inter-particle couplings.

## 3. Conclusion

In summary, from an extensive wide-angle (± 67°) *k*-space-resolved experiment on 82 antennas with different resonance conditions, with excellent agreement with the theoretical analysis, we studied the structure-dependent contributions of distinct nonlinear polarizations to far-field SHG radiation. Even for the simple geometry of a rod antenna, we show that the common belief of the surface-normal polarization dominance in SHG [13-15,27,28,35,36] could be an inaccurate assumption and should be reconsidered. Our finding demonstrates that the inclusion of all nonlinear polarization components based on a model having $|\chi_{\parallel\parallel\perp}| \sim |\chi_{\perp\perp\perp}|$ is essential for an accurate prediction of the SHG amplitude and radiation behavior in general. With the notion of a virtual multi-resonator SHG model, we then provided a method to *estimate* and *tune* individual SHG nonlinear polarization components and their far-field radiations, as an answer to the question "Is it possible to control the SHG in the far-field, by manipulating individual polarization sources?" With the increased degrees of freedom obtained from individual control of normal, parallel and bulk nonlinear polarization components, which strongly depend on the geometry and corresponding resonance condition, the results imply that the relative contribution of each source could be even more dramatic for multi-particle, metasurface arrangements with more complex geometries and layouts.

A new design strategy toward highly efficient and directive SHG devices could thus be envisaged. For example, by utilizing single particles with an intricate geometry or their combination with inter-particle couplings, SHG devices with an emphasis on a specific nonlinear polarization component could be developed. In particular, highly directional SHG emission from a bulk-polarization-dominated antenna structure or array, highly localized excitation of SHG from the use of surface-polarization-emphasized tip structures, or SHG devices based on multiple elements [27,28,30] in which the inter-particle polarization couplings are dominant could be considered for future investigations.

**Funding**

National Research Foundation of Korea (NRF) (2014M3A6B3063708, 2016R1A6A3A04009723); Brain Korea 21 Plus Project in 2019; Deutsche Forschungsgemeinschaft (SPP1839, grant LI 580/12).

**Acknowledgments**

The authors thank C. Ciracì, K. H. Lee and W. K. Park for their kind help and comments on this work.

See Supplement 1 for supporting content.

# Bridging microscopic nonlinear polarizations toward far-field second harmonic radiation: supplementary material

KYUNGWAN YOO,[1] SIMON F. BECKER,[2] MARTIN SILIES,[2] SUNKYU YU,[1,*] CHRISTOPH LIENAU,[2] AND NAMKYOO PARK[1,*]

[1]*Photonic Systems Laboratory, Department of Electrical and Computer Engineering, Seoul National University, Seoul 08826, Korea.*
[2]*Institut für Physik and Center of Interface Science, Carl von Ossietzky Universität, D-26111 Oldenburg, Germany.*
*\*Corresponding author: nkpark@snu.ac.kr, skyu.photon@gmail.com*

This document provides supplementary information to "Bridging microscopic nonlinear polarizations toward far-field second harmonic radiation," http://dx.doi.org/xx.xxxx/optica.xx.xxxxxx.s1. It includes complimentary descriptions and figures of the materials and methods, laser and SHG spectra, experimental setup, polarization dependence in second harmonic generation, antenna width dependence, calculation of nonlinear polarizations based on the hydrodynamic model, comparison with other models, dependence on the structural parameters, and SEM image for an antenna array.

## 1. Materials and methods

### 1.1 Sample fabrication

The depth of the antenna $d$ was set to be much larger (> 5 μm) than both the wavelengths of interest ($d \gg \lambda$ = 888 nm) and the incident beam waist (893 nm, full-width at half-maximum, FWHM). A high-refractive-index glass (S-LAH64) substrate was ultrasonically cleaned by using acetone, isopropyl alcohol (IPA) and deionized (DI) water with 5-min periods for each step. The cleaned substrate was spin-coated with PMMA A2 and baked on a hotplate at 175 °C for 10 min. A sacrificial polymer (Sigma Aldrich, 561223) with 1% surfactant (Sigma Aldrich, X100) was spin-coated and baked on a hotplate at 95 °C for 3 min, followed by the deposition of a 3-nm-thick anti-charging gold layer for electron beam lithography (EBL). The antennas were fabricated by an EBL system (JBX-6300FS). The patterned sample was cleaned with DI water to remove the sacrificial polymer and 3-nm anti-charging gold layer and then developed by IPA and DI water at a 3:1 ratio for 1.5 min. To pattern the antenna structure, 35 nm of gold was deposited by an electron beam evaporator, and the remaining resist was lifted off by acetone. The samples were annealed at 500 °C for a few seconds for polycrystallization [1].

### 1.2 Experimental setup and measurement of the SHG signal

The experimental setup for SHG detection and characterization is schematically shown in Fig. S2. The output beam of a mode-locked Ti:Sapphire oscillator (Venteon, Pulse: ONE), operating at 80 MHz with pulse energies of up to 2.5 nJ and a pulse duration of approximately 6 fs, was used as the excitation source. The polarization of the incident beam was controlled by a low-dispersion half-wave plate (HWP). Individual antennas of an array on the sample were placed in the focus of an all-reflective objective (36x, 0.5 NA). The residual chirp from the setup limits the pulse duration in the focal plane to 12 fs. The sample position was controlled with nanometer precision by means of a three-axis piezo stage. The nonlinear emission of the sample was collected in transmission through an immersion objective (1.65 NA, Olympus, APO 100x OHR) coupled to the sample substrate ($n$ = 1.79) with an index-matching liquid ($n$ = 1.78, Cargille, Series M). The back focal plane of the collecting objective was imaged with two off-axis parabolic mirrors in a 4-f geometry onto an electron-multiplying CCD camera. Light at the fundamental frequency was efficiently suppressed by using a combination of filters. The quadratic relation between the SHG intensity and incident power and the SHG spectrum are shown in Fig. S1.

### 1.3 Numerical simulation

We employed the approach presented in Ref. [2] to numerically obtain the *k*-space SHG data. First, a full-wave 3D antenna simulation at the fundamental frequency was carried out with a Gaussian wave excitation (FWHM = 893 nm). The results of this simulation were then used to calculate each nonlinear polarization component (surface-normal, surface-parallel and bulk) using a hydrodynamic model (see Section 2 in the Supplement 1). By using these nonlinear polarizations as a source at the second harmonic (SH) frequency, we then carried out full-wave simulations at the SH frequency. To obtain the *k*-space-resolved SHG images, the calculated SHG near fields were converted to far fields based on the Lorentz reciprocity theorem [3] with the standard apodization factor [4]. The antenna geometry used in the calculation was obtained from the TEM image of Fig. 1(a) (see Fig. S5). All the calculations were performed using commercial FEM software (COMSOL Multiphysics), while the permittivity at the fundamental and SH frequencies was taken from Olmon *et al* [5]. The implementation of the surface polarizations in the simulator was confirmed by comparing the results with analytic solutions (see Section 2 in the Supplement 1).

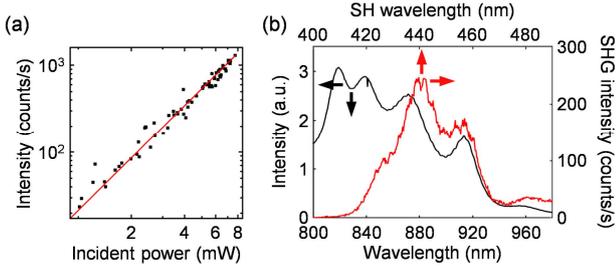

**Fig. S1.** (a) Measured SHG intensity (black symbols) as a function of the incident power on a log-log scale, well fitted by the red line with a slope of two. (b) Fundamental laser spectrum (black) and SHG spectrum (red). SHG signals below 440 nm are suppressed [6] by the absorption of the index-matching liquid ($n$ = 1.78, Cargille, Series M) used for the immersion objective (NA = 1.65, Olympus, APO 100x OHR). The SHG spectrum and laser spectrum match very well when considering the transmission curve of the index-matching liquid ($n$ = 1.78, Cargille, M series).

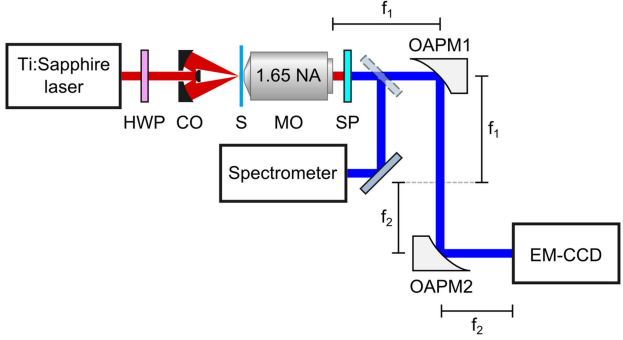

**Fig. S2.** Femtosecond pulses from a Ti:Sapphire oscillator with linear polarization controlled by a half-wave plate (HWP) are focused onto the antenna sample (S) using an all-reflective objective (CO). SHG from the sample is collected in transmission with a 1.65 NA microscope objective (MO) and passed through shortpass (SP) filters. The signal is either focused onto the entrance slit of a spectrometer or the back focal plane of the MO is imaged onto an electron-multiplying CCD camera using two off-axis parabolic mirrors (OAPM1 and OAPM2) in a 4-f geometry.

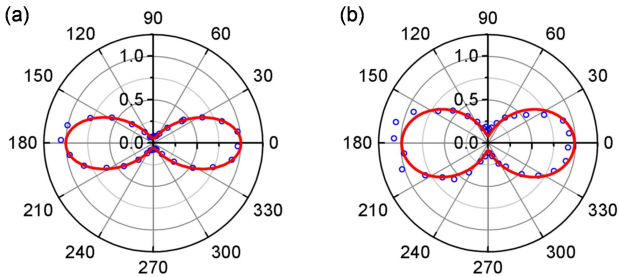

**Fig. S3.** (a),(b) Measured polarization dependence in the second harmonic generation (SHG) intensity. The antenna width is $w$ = 215 nm. In (a), the polarization angle $\phi$ of the linearly polarized fundamental incident wave is varied. The SHG radiation is then collected without conducting polarization filtering or analysis. In (b), the incident fundamental wave is linearly polarized in the $x$-direction, and the SHG is analyzed with a polarizer. The angles $\phi$ = 0° and $\phi$ = 90° correspond to the $+x$- and $+y$-directions, respectively. The measured polarization dependencies agree well with the theoretically expected $\cos^4\phi$ [red solid line in (a)] and $\cos^2\phi$ [red solid line in (b)] curves.

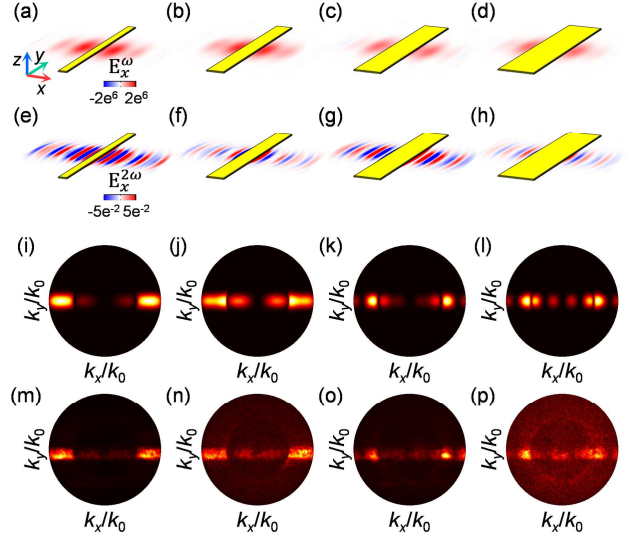

**Fig. S4.** (a)-(h) Calculated electric fields at the (a)-(d) fundamental and (e)-(h) SH frequencies. (i)-(l) Calculated and (m)-(p) measured SHG intensities in the $k$-space plane. Each column shows the results for antenna widths of $w$ = 205, 420, 645, and 870 nm, from left to right.

## 2. Calculation of nonlinear polarizations based on the hydrodynamic model

In the hydrodynamic model (HDM) [7], the optical response of metals is described in terms of the collective motion of the free electrons inside the metal. The equations of motions in the HDM are formulated for the charge density and velocity of the electron fluid under the influence of the Lorentz force, quantum pressure and convective acceleration. This model can describe several features that are commonly not included in the macroscopic Maxwell's equation, such as longitudinal modes [8], an insensitivity to small-scale roughness [9], or second-, third- and higher-harmonic generation processes [10-12]. Specifically, the HDM inherently includes a nonlinear term that can be expanded in a perturbation series in relation to inducing second- and third-harmonic radiation.

At the early stage of an HDM-based SHG study, Sipe *et al.* [13] introduced an effective sheet, which allows the explicit expression of the surface-parallel ($P^{2\omega}_{\text{surf},\parallel}$) second-order nonlinear polarization. Further development was made by Ciracì *et al.* [2] in deriving an explicit expression of the surface-normal ($P^{2\omega}_{\text{surf},\perp}$) component in addition to $P^{2\omega}_{\text{surf},\parallel}$. Here, we follow Ref. [2], which provides the best fit to our experimental data (see Fig. 2 and Section 3 in the Supplement 1 for a detailed comparison). Each component of the nonlinear polarizations is then expressed as:

$$P^{2\omega}_{\text{surf},\perp} = \varepsilon_0 \chi_{\perp\perp\perp} E^{\omega}_{\perp} E^{\omega}_{\perp}, \quad \text{(S1)}$$

$$P^{2\omega}_{\text{surf},\parallel} = \varepsilon_0 \chi_{\parallel\parallel\perp} E^{\omega}_{\parallel} E^{\omega}_{\perp}, \quad \text{(S2)}$$

$$\mathbf{P}^{2\omega}_{\text{bulk}} = \varepsilon_0 \alpha \mathbf{E}^{\omega} \times \mathbf{B}^{\omega} + \varepsilon_0 \beta'(\mathbf{E}^{\omega} \cdot \nabla)\mathbf{E}^{\omega}, \quad \text{(S3)}$$

where

$$\chi_{\perp\perp\perp} = \frac{e}{4m_e^*}\frac{3\omega - i\nu}{2\omega - i\nu}\frac{1}{\omega^2 - i\omega\nu}\chi^{\omega}, \quad \text{(S4)}$$

$$\chi_{\parallel\parallel\perp} = \frac{e}{2m_e^*}\frac{1}{\omega^2 - i\omega\nu}\chi^{\omega}, \quad \text{(S5)}$$

$$\alpha = \frac{e}{m_e^*} \frac{i\omega}{4\omega^2 - 2i\omega\nu} \chi^\omega, \quad (S6)$$

$$\beta' = -\frac{e}{m_e^*} \frac{1}{4\omega^2 - 2i\omega\nu} \frac{1}{\omega^2 - i\omega\nu} \chi^\omega, \quad (S7)$$

$$\chi^\omega = \frac{\omega_p^2}{\omega^2 - i\omega\nu}, \quad (S8)$$

$$\omega_p^2 = \frac{n_0 e^2}{m_e^* \varepsilon_0}. \quad (S9)$$

Here, $m_e^*$ is the effective electron mass, $n_0$ is the background electron charge density, $\nu$ is the electron collision rate, $-e$ is the charge of an electron. We used the following values for all the calculations: $m_e^* = 1.06 \times m_e^*$, $n_0 = 5.9 \times 10^{22}$ cm$^{-3}$, and $\nu = 1.07 \times 10^{14}$ s$^{-1}$. The superscripts $\omega$ and $2\omega$ denote the values at the fundamental ($1\omega$) and SH ($2\omega$) frequencies, respectively, and the subscripts $\perp$ and $\parallel$ denote the surface-normal and surface-parallel components, respectively. In deriving Eqs. (S1)-(S9), a harmonic time dependence [$\exp(i\omega t)$] has been assumed.

It is also noted that from the Maxwell's equation and the vector identity, the expression for the bulk nonlinear polarization can be written as [14]:

$$\mathbf{P}_{bulk}^{2\omega} = \varepsilon_0 \gamma \nabla(\mathbf{E}^\omega \cdot \mathbf{E}^\omega) + \varepsilon_0 \delta'(\mathbf{E}^\omega \cdot \nabla)\mathbf{E}^\omega, \quad (S10)$$

where

$$\gamma = -\frac{e}{2m_e^*} \frac{1}{4\omega^2 - 2i\omega\nu} \chi^\omega, \quad (S11)$$

$$\delta' = -\frac{e}{m_e^*} \frac{1}{4\omega^2 - 2i\omega\nu} \frac{i\omega\nu}{\omega^2 - i\omega\nu} \chi^\omega. \quad (S12)$$

We utilized the commercial software COMSOL Multiphysics for the calculation of the electric fields at the fundamental frequency and its application to the modeling of the nonlinear polarization sheets (see Section 1 for details). The validity of the COMSOL calculation was confirmed by comparing the results with the analytical method [15,16].

## 3. Comparison with other models

The second-order nonlinear susceptibility tensor $\chi^{(2)}$ (e.g., $\chi_{\perp\perp\perp}$, $\chi_{\parallel\parallel\perp}$, $\chi_{\perp\parallel\parallel}$, $\gamma$, and $\delta'$) has been intensively studied in different experimental and theoretical settings, including reflections from a flat metal film [17,18], spherical structures [19,20], two-beam excitation [21], a free electron model [22] and a hydrodynamic model [2,13]. Here, we compare the SHG simulations with our experimental result by taking $\chi^{(2)}$ parameters from various reports in the literature [2,17,19,21,22]. To compare $\chi^{(2)}$ parameters, we convert the $\chi^{(2)}$ parameters to conventional notations [21], where the surface nonlinear polarizations are calculated by evaluating the fundamental fields just inside the metal, and placed just outside the metal. The different definitions for $\chi_{\parallel\parallel\perp}$ in relation with redundant terms [18] are also considered. The wavelength dependence of $\chi^{(2)}$ is considered by adopting the model in [23] and [19] as

$$\chi_{\perp\perp\perp} = -\frac{ea}{4m_e^* \omega^2} \chi^\omega, \quad (S13)$$

$$\chi_{\parallel\parallel\perp} = -\frac{ea}{2m_e^* \omega^2} \chi^\omega, \quad (S14)$$

$$\gamma = -\frac{ea}{8m_e^* \omega^2} \chi^\omega, \quad (S15)$$

where $a$, $b$, and $d$ are Rudnick and Stern (RS) parameters [23], which are treated as frequency-independent in the case of $\omega \ll \omega_p$ [24,25]. Table S1 shows the converted $\chi^{(2)}$ parameters using Eqs. (S13)-(S15) at a wavelength of 888 nm.

It is noted that in Wang et al. [21], bulk parameter $\gamma$ was presented in the form of effective surface parameters, through $\chi_{\perp\perp\perp}^{\text{eff}} = \chi_{\perp\perp\perp} + \gamma/\varepsilon(2\omega)$ and $\chi_{\perp\parallel\parallel}^{\text{eff}} = \chi_{\perp\parallel\parallel} + \gamma/\varepsilon(2\omega)$. However, considering that $\chi_{\perp\parallel\parallel}$ is predicted to be zero in theoretical point of view [17], Wang's $\chi_{\perp\perp\perp}^{\text{eff}}$ and $\chi_{\perp\parallel\parallel}^{\text{eff}}$ reduces to $\chi_{\perp\perp\perp}$ and $\gamma/\varepsilon(2\omega)$, where $\chi_{\perp\perp\perp}$ is still an order of magnitude larger than $\gamma/\varepsilon(2\omega)$ and also for all the other parameters. This indicates that the ambiguity for $\gamma$ parameter classification (as bulk term or surface term) does not affect our results, as confirmed in our numerical analysis testing both cases (not shown here). Finally, the contribution of bulk parameter $\delta'$ was found to be minor as well (not shown here), in good agreement with previous report [21].

**Table S1. Second-order nonlinear susceptibility tensor components for gold at 888 nm.**

| | Second-order susceptibility tensor components at 888 nm (m$^2$/V) | | | | | Relative magnitudes of the RS parameter | | | Relative phases of the RS parameter (°) | | | Ref. |
|---|---|---|---|---|---|---|---|---|---|---|---|---|
| | $\chi_{\perp\perp\perp}$ | $\chi_{\parallel\parallel\perp}$ | $\gamma$ | $\delta'$ | $\chi_{\perp\parallel\parallel}$ | $|a|$ | $|b|$ | $|d|$ | $\angle a$ | $\angle b$ | $\angle d$ | |
| Krause et al. | -4.4e-18 -3.0e-18i | -2.8e-19 +2.5e-20i | 1.1e-20 +4.8e-23i | - | - | 14 | 0.4 | 0.06 | 211 | 172 | -3 | [17] |
| *Wang et al. | 7.6e-18 -7.5e-20i | 1.1e-18 -1.1e-20i | [-1.4e-19 -7.2e-20i] | 3.6e-19 -8.9e-21i | 3.0e-20 -3.0e-22i | 20 | 1.5 | [0.8] | - | - | - | [21] |
| Bachelier et al. | 2.1e-19 -7.9e-20i | 8.2e-20 +5.3e-21i | 1.9e-19 +1.2e-20i | - | - | 0.59 | 0.11 | 0.97 | -23 | 1 | 1 | [19] |
| Benedetti et al. | 4.8e-18 +9.4e-19i | -8.1e-19 -7.7e-20i | 2.0e-19 +1.2e-20i | -1.5e-21 +1.5e-20i | - | 13 | 1.1 | 1.1 | 8 | 183 | 0 | [22] |
| Ciracì et al. | -5.4e-19 -5.9e-20i | -7.2e-19 -7.3e-20i | 1.8e-19 +1.4e-20i | -2.3e-21 +1.8e-20i | - | 1.4 | 0.9 | 0.9 | 183 | 183 | 1 | [2] |

*These parameters are calculated from those conventional definition values listed in Wang [21]. The parameter $\gamma$ in brackets is listed with the assumption of $\chi_{\perp\parallel\parallel}^{\text{eff}} \sim \gamma/\varepsilon(2\omega)$ [17], for comparison purpose only (not used in the calculation). The same assumption $\chi_{\perp\parallel\parallel}^{\text{eff}} \sim \gamma/\varepsilon(2\omega)$ [17] is used for the wavelength conversion of $\chi_{\perp\parallel\parallel}^{\text{eff}}$ with Eq. (S15). For the wavelength conversion of Wang's $\delta'$ value, HDM interpretation of $\delta'$ [Eq. (S12)] is used.

## 4. Dependence on the structural parameters

In this section, we examine the dependence of SHG radiation on the structural parameters of the antenna. We classify the antenna geometry with 3 geometric parameters based on the TEM image [Fig. 1(a) in the main manuscript]: the rounding radius of the bottom edge ($r$), width difference ($diff$) between the top and bottom surfaces, and thickness ($t$) of the antenna [Fig. S5(a)]. Figures S5(b)-S5(g) show the dependence of the strengths of the electric fields on the structural parameters at the fundamental [Figs. S5(b)-S5(d)] and second-harmonic [Figs. S5(e)-S5(g)] frequencies. For the sharper (smaller radius) bottom edge ($r$), the SHG intensities from $P^{2\omega}_{surf,\perp}$ and $P^{2\omega}_{surf,\parallel}$ are enhanced [Fig. S5(e)], following $E^{\omega}_{\perp}$ of the fundamental electric field [Fig. S5(b)]. On the other hand, for an antenna with a smaller thickness ($t$), the SHG intensities of $P^{2\omega}_{surf,\parallel}$ and $\mathbf{P}^{2\omega}_{bulk}$ are enhanced [Fig. S5(g)], following $E^{\omega}_{\parallel}$ of the fundamental electric field [Fig. S5(d)]. Figures S5(c) and S5(f) indicate that the resonance of the antenna is predominantly governed by the length at the bottom surface, not by the total length of the antenna. These results show the possibility of independent control of each nonlinear polarization by combining structural modifications.

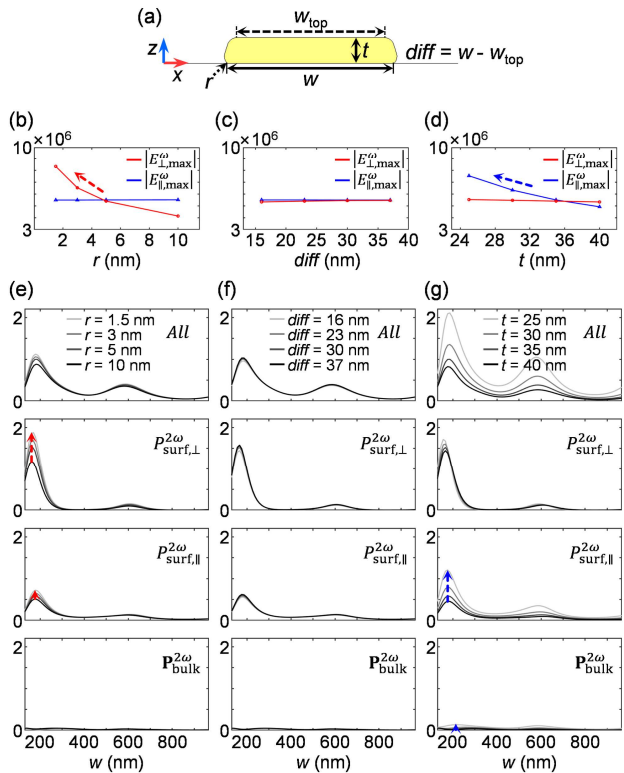

**Fig. S5.** (a) Geometry of the antenna. (b)-(d) Maximum strength of the fundamental electric field $E^{\omega}_{\perp}$ and $E^{\omega}_{\parallel}$, for $w$ = 205 nm, plotted as a function of the (b) rounding radius of the bottom edge, (c) width difference between the top and bottom surface, and (d) thickness. (e)-(g) Integrated transmitted SHG intensity as a function of antenna width, with different structural variations. Each row of (e)-(g) shows the results from all nonlinear polarizations, surface-normal, surface-parallel, and bulk nonlinear polarizations, from top to bottom.

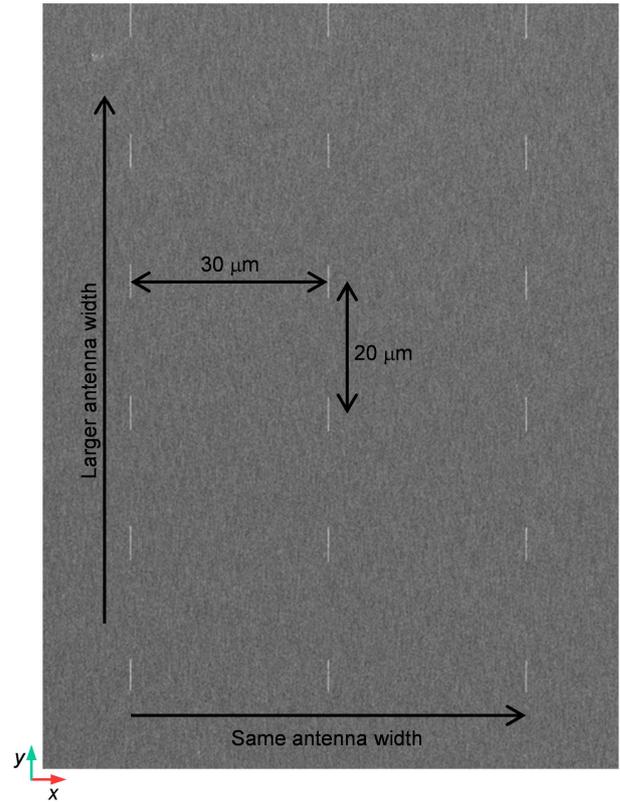

**Fig. S6.** Antenna array consisting of antennas with different widths. The space between adjacent antennas was set large enough to avoid nearest-neighbor coupling both at the fundamental and the second-harmonic frequencies.